# Superconducting wide strip photon detector with high critical current bank structure

Masahiro Yabuno[1,*], Fumihiro China[1], Hirotaka Terai[1], and Shigehito Miki[1,2]

[1.] Advanced ICT Research Institute, National Institute of Information and Communications Technology, 588-2 Iwaoka, Nishi-ku, Kobe 651-2492, Japan
[2.] Graduate School of Engineering, Kobe University, 1-1 Rokkodai-cho, Nada, Kobe 657-0013, Japan

* masahiro.yabuno@nict.go.jp

**Abstract**

Superconducting strip single-photon detectors offer excellent photon detection performance and are indispensable tools for cutting-edge optical science and technologies, including photonic quantum computation and quantum networks. Ultra-wide superconducting strips with widths of tens of micrometers are desirable to achieve high polarization-independent detection efficiency using a simple straight strip. However, biasing the ultra-wide strip with sufficient superconducting current to make it sensitive to infrared photons is challenging. The main difficulty is maldistribution of the superconducting current in the strip, which generates excessive intrinsic dark counts. Here, we present a novel superconducting wide strip photon detector (SWSPD) with a high critical current bank (HCCB) structure. This HCCB structure enables suppression of the intrinsic dark counts and sufficient superconducting current biasing of the wide strip. We have experimentally demonstrated a polarization-independent system detection efficiency of ~78% for 1550 nm wavelength photons and a system dark count rate of ~80 cps using a 20-μm-wide SWSPD with the HCCB structure. Additionally, fast jitter of 29.8 ps was achieved. The photolithographically manufacturable ultra-wide SWSPD with high efficiency, low dark count, and fast temporal resolution paves the way toward the development of large-scale optical quantum technologies, which will require enormous numbers of ultimate-performance single-photon detectors.

**Main text**

Single-photon detectors are essential for a wide range of scientific research applications, from bioscience to astrophysics. Emerging quantum technologies such as quantum networks and quantum computers will require enormous numbers of ultimate-performance single-photon detectors.

Although various detectors, including photomultiplier tubes, avalanche photodiodes, and superconducting detectors, have been developed to detect single photons efficiently, a single-photon detector with a current-carrying superconducting strip is one of the most promising candidate devices. Superconducting nanostrip single-photon detectors (SNSPDs or SSPDs), which use superconducting strips with widths of approximately 100 nm, were first demonstrated in 2001, and these detectors have seen rapid growth over the past two decades [1–2]. Excellent detection performances have been achieved, including system detection efficiency (SDE) of more than 98% [3,4], a low dark count rate (DCR) of less than 0.0001 Hz [5], fast timing jitter of less than 3 ps [6], and a high maximum counting rate exceeding 1 GHz [7]. Moreover, single-photon imaging [8–15] and photon-number-resolving detection [16–21] have been demonstrated using arrayed SNSPDs and/or specific readout techniques. These features have made SNSPDs essential devices not only for quantum technologies [22–26], but also for a broad range of applications from bioimaging [27,28] to remote sensing [29–31] and space communications [32].

In recent years, superconducting microstrip single-photon detectors (SMSPDs) with strip widths of 1 μm or more have offered new potential for development of superconducting strip single-photon detectors. For many years, narrow superconducting strips with widths of around 100 nm have been considered essential for single-photon detection, particularly in the infrared wavelength range. In 2018, Korneeva *et al*. demonstrated near-infrared single-photon detection using superconducting strips with widths of a few micrometers [33], with experimental results that were consistent with predictions based on the vortex model [34]. Their innovative research has excited interest in the physics inside the current-carrying superconducting strip and has raised a fundamental question about the upper limit to the strip width that allows a single photon to be detected. Wider superconducting strips with micrometer-scale widths offer many practical advantages that the nanometer-scale strips cannot provide. A superconducting strip that is wider than the optical spot diameter will enable realization of a simple straight strip detector that has high efficiency because of its 100% filling factor and polarization insensitivity because of the unpatterned geometric shape of its photo-receiving area. Wider strips that can be fabricated via optical lithography will improve both fabrication throughput and yield, which will increase the number of pixels contained in single-photon imagers and photon-number-resolving detectors. In addition, the wider strips will be advantageous in allowing construction of a larger photo-receiving area while maintaining low kinetic inductance and a high temporal response. To realize these advantages, SMSPDs based on various superconducting materials have been developed actively over the past few years [35–43]. In addition, the geometries of these superconducting strips have been studied to reduce the geometrical current-crowding effect, which reduces the switching bias current and limits detection efficiency [44–47]. As a result of these challenges, SDEs of more than 90% have been achieved using superconducting strips with widths in the vicinity of 1 μm [46,47]. However, despite the practical importance, detection of single photons using broader superconducting strips with widths of more than 10 μm while achieving high detection efficiency remains highly challenging. One of the main causes of this difficulty is the detector's excessive intrinsic dark counts, which increase with increasing strip width and bias current.

To reduce these intrinsic dark counts, we propose a novel superconducting strip structure that has two different critical currents across the strip, with the edge regions having higher critical currents than the center region. We call this structure the high critical current bank (HCCB) structure. Using this structure, we demonstrate single-photon detection using a 20-μm-wide niobium-titanium nitride (NbTiN) superconducting wide strip photon detector (SWSPD) operating in the near-infrared wavelength range. The proposed strip structure successfully suppresses increases in the intrinsic dark counts while also supplying increased bias current to the wide strip. As a result, high SDE is achieved together with a low DCR. In addition, we also demonstrate the polarization insensitivity and fast timing jitter of the SWSPD under high bias current conditions. These results represent a major step toward ultra-wide SWSPDs with high SDE, low DCR, and fast timing jitter. Furthermore, these results will provide valuable insights for investigations into the dynamics of photon detection and dark count generation in current-carrying superconducting strips.

**Concept of the SWSPD with the HCCB structure**

High-efficiency single-photon detection using wider superconducting strips requires a sufficiently high bias current to be supplied to the strips while the intrinsic dark count must be suppressed. However, in practice, the distribution of the bias currents is concentrated near the edges of the strip because of the influence of the magnetic flux density distribution [48]. Because SNSPDs and SMSPDs are usually designed to have uniform thickness and superconductivity across the widths of their strips, the concentration of the bias current induces vortex penetration from the side edges that leads to a rapid increase in the intrinsic dark counts with increasing bias current. As a result, the strip switches permanently to the normal conducting state before a sufficient bias current can be supplied to the central region of the strip.

To solve this problem, we propose a superconducting strip with an HCCB structure, as drawn schematically in Fig. 1. In the proposed device, both of the strip's side edge regions are structured to have higher critical currents per unit width than the central region. These side strip regions suppress vortex penetration into the strip, thus reducing the intrinsic dark count. As a result, sufficiently high bias current can be applied more evenly to the central region along the strip width. In addition, because the central area acts as the photon detection area and the side regions are not involved, the sensitivity limitations caused by defects on the side edges of the strip, e.g., as caused by the manufacturing process, can be suppressed to some extent.

**Detection performance of conventional SMSPD with wide width**

Before testing our device concept, we investigated the photon detection characteristics of an SMSPD that was designed to have uniform superconductivity across the width of its strip. Figure 2 shows the bias current dependences of the SDE and DCR of the approximately 4.9-nm-thick NbTiN straight strip detector with a strip width of 20 μm and length of 20 μm measured at temperatures of (a) 0.76 K and (b) 2.2K. The detector shape was the same as that used for the novel detector concept, which will be described later, except for the structure of the side strip areas. The wavelength of the photons irradiating the detector was 1550 nm, and the average photon number was tuned to $1 \times 10^5$ photons/s. The 20-μm-wide SMSPD device detected single photons at a near-infrared wavelength successfully. However, as mentioned in the concept described above, the rapid increase in the intrinsic dark count limited device operation in the high bias current region. As a result, the device was permanently switched to the normal conducting state before reaching SDE saturation. Specifically, at a temperature of 2.2 K, the intrinsic dark counts rose from much lower bias currents, causing device operation in the high bias current region to be strongly limited. Therefore, suppression of the intrinsic dark counts is a significant requirement to produce higher detection efficiency when using the wider superconducting strip.

**Fabrication process of the SWSPD with the HCCB structure**

In this work, we realized the proposed strip structure by using argon (Ar) ion beam irradiation to reduce the critical current per unit strip width in the central region. Ar ion beam irradiation effectively reduces the critical temperature of the thin superconducting strip and the maximum superconducting current that can be applied to that strip. For example, the critical temperature of the 10-μm-wide NbTiN strip with thickness of approximately 5.9 nm was degraded from 8.2 K to 6.4 K after 5 s of irradiation with the Ar ion beam (see Extended Data Fig. 2 for details of the preliminary experiments). Figure 3 shows schematics and micrographs of SWSPD devices that were irradiated using the Ar ion beam. We fabricated two SWSPD device types for comparison that were irradiated with Ar ion beams over different areas on their strips. Figure 3a shows the SWSPD device that was irradiated by the Ar ion beam over the central area of the strip to form the side strip areas with higher critical currents. The area irradiated by the Ar ion beam is bounded by the red dashed line in the micrograph. Figure 3b shows that the SWSPD device that was irradiated with the Ar ion beam over the entire width of the strip has uniform superconductivity across the strip width. Note that the devices shown in Fig. 3 do not have optical cavity structures to increase photon absorption. To fabricate these devices, a NbTiN thin film with thickness of approximately 5.9 nm was first deposited on a silicon wafer with a thermally oxidized layer using DC magnetron reactive sputtering. The NbTiN film was then patterned to form a wide strip with a length of 20 μm, a width of 20 μm, and a coplanar waveguide (CPW) structure using maskless photolithography and reactive ion etching. Subsequently, the Ar ion beam was used to irradiate the wide strip over a length of 15 μm to reduce the critical currents in the strips. For the SWSPD device with the HCCB structure, the width of the area irradiated by the Ar ion beam was 18 μm, meaning that the width of the side strip area was 1 μm for each side region. The fabricated devices were cooled down to the 0.76 K to 2.2 K temperature range using a He-4 sorption refrigerator, and the detection performance for the 1550 nm wavelength photons was then evaluated.

**Detection performance of the SWSPD with the HCCB structure**

Figure 4 shows the bias current dependences of the SDE and the DCR of the two types of SWSPD device at temperatures of 0.76 K and 2.2 K. The photons emitted from a continuous wave (CW) laser source were attenuated to an intensity of $1\times10^5$ photons/s and propagated into a refrigerator via a single-mode optical fiber after passing through a polarization control module. The spot size on the wide strip caused by photon illumination from the back side of the device was tuned to a $1/e^2$ diameter of approximately 8.2 μm, and the spot center was aligned with the center of the photo-receiving area. To prevent a permanent transition into the normal conducting state and to ensure stable operation of the SWSPD devices, a series-connected 220 nH inductor and a parallel-connected 12.5 Ω shunt resistor were used on the outside of the radio-frequency (RF) input/output port of the refrigerator (see the Methods and Extended Data Fig. 1 for details of the experimental setup).

The logarithmically displayed dark count curves have two regions that have different slopes. The counts that increase slowly toward saturation in the low bias current range are caused by the extrinsic dark count, which is mainly blackbody radiation, while the counts that increase steeply within the high bias current range are caused by the intrinsic dark count that is inherent to these devices. The intrinsic dark count of the SWSPD without the side strip areas increased from a relatively low bias current, resulting in the device being switched into the normal conducting state before its detection efficiency approached saturation. In contrast, the SWSPD with the HCCB structure suppressed the rise of the intrinsic dark count to the higher bias current range successfully and had a longer extrinsic dark count region. As a result, the detection efficiency of the device was clearly saturated with a broad plateau region, indicating that the internal detection efficiency reached 100%. These results clearly demonstrate the effect of the HCCB structure in suppressing the intrinsic dark count and thus increasing the maximum bias current that can be supplied to the wide strip, resulting in improved detection efficiency.

Next, to improve the SDE, the SWSPD with the HCCB structure was integrated into a double-sided optical cavity [49] designed to enhance the optical absorption at a wavelength of 1550 nm. The NbTiN film thickness was reduced slightly to 5.6 nm in expectation of it having increased sensitivity to near-infrared photons. Figure 5 shows the bias current dependences of the SDE and the DCR of the SWSPD with the HCCB structure when integrated inside the optical cavity. The inset shows a schematic of the cross-sectional structure of the SWSPD with the double-sided optical cavity. The SDE of ~78% with a broad plateau region and the DCR of ~80 cps were achieved at a temperature of 0.76 K. Because the SDE curve saturated at bias currents that were sufficiently lower than the intrinsic dark count rises, the system dark counts could be reduced further by reducing the extrinsic factor via appropriate filtering, e.g., using a cold optical filter [50]. In addition, the broad plateau region in the SDE curve indicates that the detector has high sensitivity, even for photons at much longer wavelengths.

When the temperature increases, the intrinsic dark count rises at lower bias currents, thus limiting the maximum bias current at which the device can operate stably. However, even at a temperature of 2.2 K, the detector approaching saturation of the SDE curve achieved SDE of ~76% and a DCR of less than 100 cps.

Figure 6a shows the SDE's wavelength dependence within the range from 1548 to 1552 nm when measured at a bias current of 2.7 mA and a temperature of 0.76 K. The variation in the SDE with a period of approximately 0.7 nm is due to optical interference induced between the back and front surfaces of the silicon wafer with the thermally oxidized layer. At wavelengths within the vicinity of the top of the variation curve, SDEs of ~80% were achieved. We believe that further improvements in the SDE can be achieved via appropriate fine tuning of the optical cavity design and fabrication process.

Figure 6b shows the SDE measured at a bias current of 2.7 mA and a temperature of 0.76 K while the polarization state of the photon input to the refrigerator system was varied. The polarization state of the photon was manipulated using a polarization control module consisting of a polarizer, a half-wave plate, and a quarter-wave plate. The SDE varied from a maximum of 78.7% to a minimum of 78.0% as the result of rotation of the half-wave plate from 0 to 90° and the quarter-wave plate from 0 to 45° in 5° steps, respectively. These results indicate that the detector has quite low polarization sensitivity.

Finally, we evaluated the timing jitter performance of the SWSPD with the HCCB structure using a femtosecond pulsed laser operating at a wavelength of 1550 nm with a pulse duration of 100 fs and a 10 MHz repetition rate. The number of photons per pulse was set at 0.001 photons/pulse. Figure 6c shows the timing jitter histogram when measured at a bias current of 2.8 mA and a temperature of 0.76 K. The full width at half maximum timing jitter of 29.8 ps was achieved using the 20-μm-wide SWSPD with the HCCB structure.

**Conclusion**

We have demonstrated a novel superconducting wide strip detector with side strip areas that have higher critical currents per unit width than the central area of the wide strip. Using this HCCB structure, the intrinsic dark count was suppressed effectively even in photon detectors with wider superconducting strips. As a result, by supplying sufficiently high bias currents, SDE of ~78% and a DCR of ~80 cps were achieved using the SWSPD with the HCCB structure, which has a total strip width of 20 μm and is integrated into an optical cavity that is designed to enhance optical absorption at a wavelength of 1550 nm. The bias current dependence of the SDE showed a broad plateau region, indicating that the internal detection efficiency reached 100% at the 1550 nm wavelength and that the detector has high sensitivity, even for photons with wavelengths longer than 1550 nm. Because the strip width was sufficiently wider than the optical spot diameter, the detection efficiency was independent of the polarization state of the input photons. The SDE could be improved further by optimizing the optical cavity design and fabrication process. The system's dark count could also be reduced further by reducing the extrinsic factor via appropriate filtering. We also demonstrated timing jitter of 29.8 ps using the 20-μm-wide SWSPD with the HCCB structure. This result suggests that by applying a sufficiently high bias current, fast timing jitter could be achieved even for photon detectors with wider superconducting strips. These results open up the possibility of development of ultra-wide SWSPDs with high detection efficiencies, low dark counts, and fast temporal resolution, and will help to reveal the origin of the intrinsic dark count in current-carrying superconducting strips. These ultra-high performance SWSPDs, which can be mass-fabricated with high yields via photolithography, would accelerate the realization of future large-scale quantum information and communication technologies, which will require enormous numbers of ultimate-performance single-photon detectors.

## Methods

### Device fabrication

First, a niobium titanium nitride (NbTiN) thin film was deposited on a silicon (Si) wafer with a thermally oxidized silicon dioxide ($SiO_2$) layer using reactive DC magnetron sputtering. The Si wafer thickness was 450 μm, including 260-nm-thick $SiO_2$ layers on both surfaces. The NbTiN thin film thickness was controlled via the deposition time and tuned to a range around 4.9 to 5.9 nm. The NbTiN thin film was then processed into wide strips with a coplanar waveguide structure using photolithography and reactive ion etching processes. In this research, a maskless aligner (MLA 150, Heidelberg Instruments GmbH) and a positive photoresist (AZ MiR 703) were used in all the photolithography processes. To modulate the superconductivity of the NbTiN thin film and form the photo-receiving area, an Ar ion beam was used to irradiate the wide strips, which were masked with a patterned photoresist layer, for 5 s. The beam voltage and current were adjusted to 300 V and 250 mA, respectively. An optical cavity structure to enhance photon absorption was formed using a lift-off process. First, a silicon monoxide (SiO) insulation layer with a thickness of 204 nm was deposited on the wide strip via thermal vapor deposition. A 150-nm-thick gold (Au) mirror with a 3-

nm-thick titanium (Ti) adhesion layer was then deposited by thermal vapor deposition. Finally, these three layers were lifted off and a square-shaped optical cavity structure was formed on the photo-receiving area of the wide strip.

*Detection efficiency measurement*

The detector was cooled using a He-4 sorption refrigerator with a lowest attainable temperature of 0.76 K. A tunable CW laser source (TSL-550, Santec) with a tunable wavelength range from 1480 to 1630 nm was used as the photon source. The input photon number was controlled to $1\times10^5$ photons/s on average using an optical power controller (81576A, Keysight Technologies) and a variable optical attenuator (81570A, Keysight Technologies). The polarization state of the input photons was controlled using a manually controlled polarization controller (Optoquest) that consisted of a polarizer, a half-wave plate, and a quarter-wave plate. The photons were introduced into a refrigerator via a single-mode optical fiber (SMF-28e) and focused on the photo-receiving area on the wide strip from the back side of the device chip using a gradient-index (GRIN) lens fabricated on the tip of the optical fiber. The working distance of the GRIN lens was 150 μm and the spot size on the wide strip was approximately 8.2 μm ($1/e^2$ diameter). The bias current for the detector was supplied through a bias tee (5575A, Picosecond Pulse Labs) from a voltage source (SIM928, Stanford Research Systems) with a 1 kΩ resistor connected in series. At the RF input/output port on the exterior of the refrigerator, a 220 nH inductor was connected in series and a 12.5 Ω shunt resistor was connected in parallel. These electrical components prevented latching and stabilized detector operation. An output signal from the detector was amplified using a low-noise amplifier (LNA-1800, RF Bay) and counted using a photon counter (SR400, Stanford Research Systems).

*Timing jitter measurement*

A 1550 nm femtosecond-pulsed laser source (FPL-02CFF, Calmar Laser) with pulse duration of ~100 fs and a repetition rate of 10 MHz was used as the photon source in place of the CW laser source. All other optical components and their arrangements were the same as those used for the detection efficiency measurement. The number of photons per pulse was controlled to be 0.001 photons/pulse on average. The electrical components used to drive the detector and the low-noise amplifier for output signal amplification were also the same components that were used to perform the detection efficiency measurements. The output signals from the detector and the synchronization signals from the laser source were counted using a time-correlated single photon counter (HydraHarp 400, PicoQuant GmbH) to measure the timing histogram of the detected events.

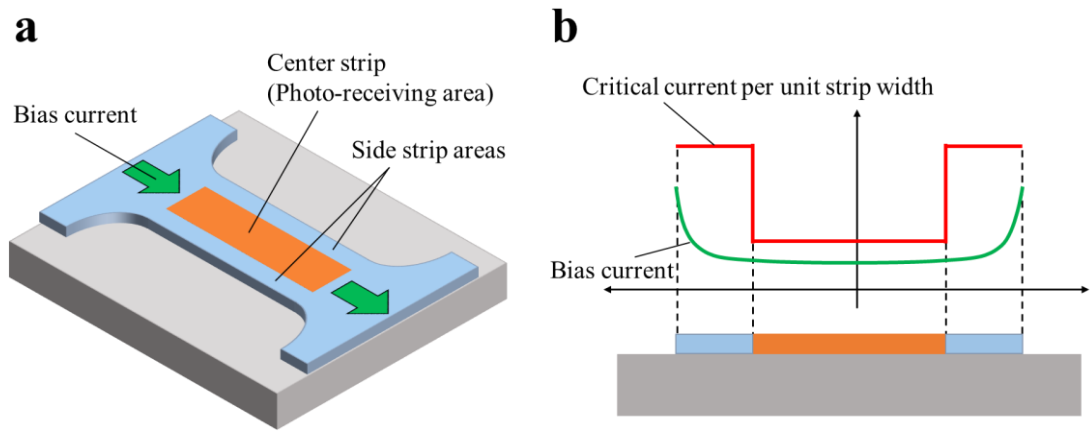

**Fig. 1 | Concept of the novel SWSPD with the HCCB structure. a**, Schematic of the SWSPD device with a center strip area acting as the photo-receiving area and side strip areas formed outside the center strip area. **b**, Cross-sectional view of the SWSPD device and schematic image showing the distribution of the critical current per unit strip width of the wide strip and the bias current flowing in the wide strip. The bias current flowing in the superconducting strip is concentrated near the edges, which induces vortex penetration from the side edges and leads to the rapid increase in the intrinsic dark counts. The novel superconducting strip structure has side strip areas that have higher critical currents per unit strip width than the center strip area to suppress both vortex penetration into the strip and the increase in the intrinsic dark count. We call this structure the high critical current bank (HCCB) structure.

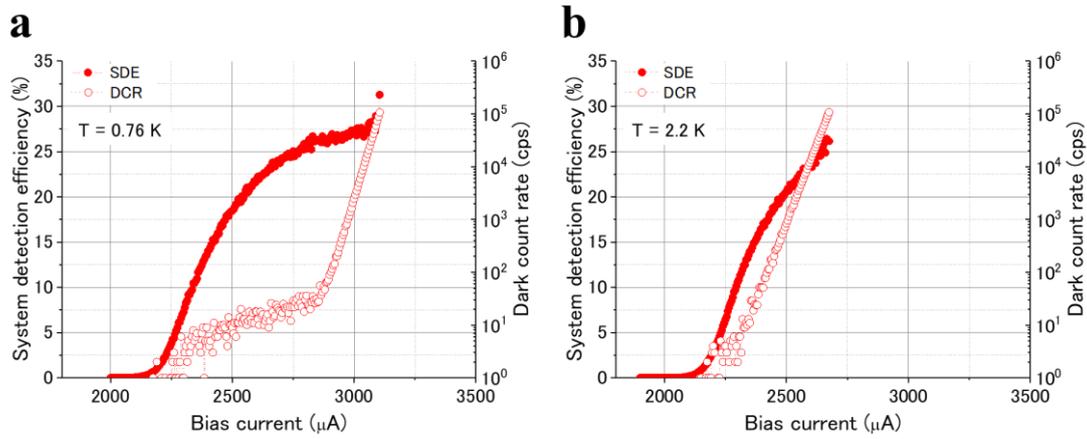

**Fig. 2 | Bias current dependences of the SDE and DCR of the conventional SMSPD with broad width. a**, SDE (red filled circles) and DCR (red open circles) curves measured at a temperature of 0.76 K. The NbTiN microstrip thickness was approximately 4.9 nm. The strip width and length were both 20 μm. The photon wavelength at which the strip was irradiated was 1550 nm. The rapid increase in the intrinsic dark count prevented device operation before the SDE curve reached a plateau. **b**, SDE and DCR curves when measured at a temperature of 2.2 K. The intrinsic dark count increased from much lower bias currents and limited device operation severely.

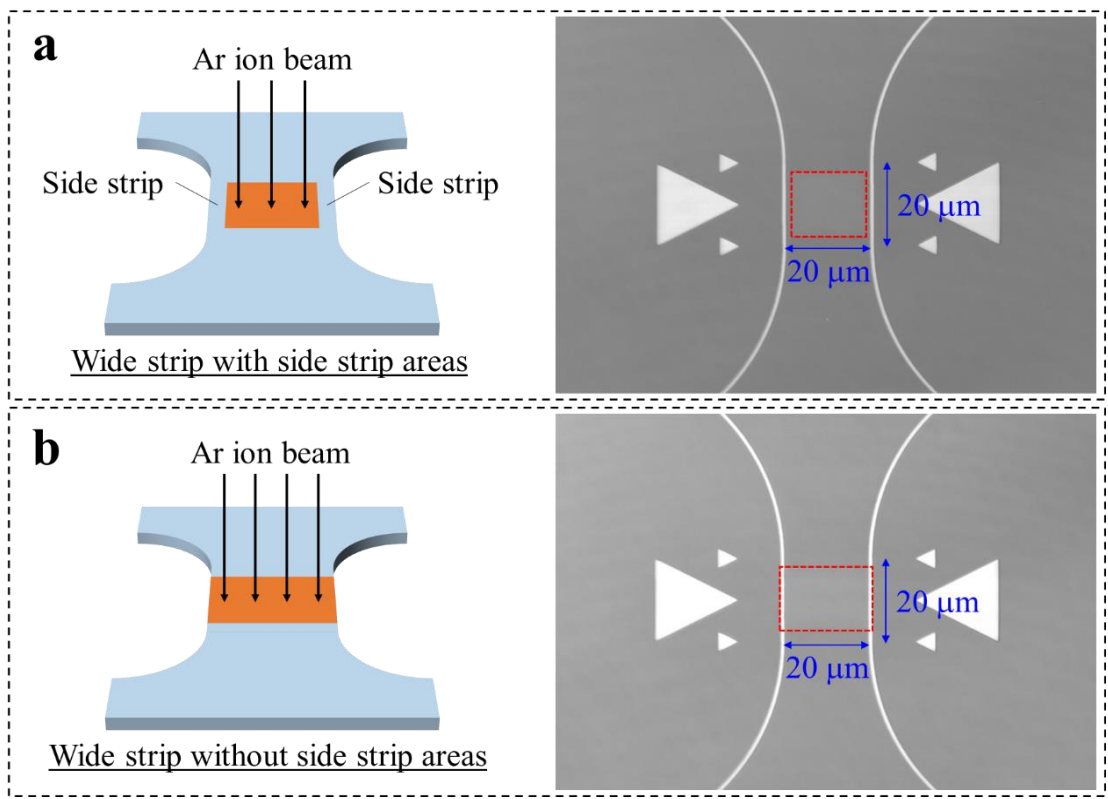

**Fig. 3 | Schematics and micrographs of the SWSPD devices that were irradiated using the Ar ion beam. a**, SWSPD device with the HCCB structure. The Ar ion beam irradiated the central area of the wide strip to form side strip areas with high critical current per unit strip width. The area irradiated by the Ar ion beam is bounded by the red dashed line shown in the micrograph. **b**, SWSPD device without the HCCB structure. The Ar ion beam irradiated the entire width of the wide strip, causing the wide strip to have uniform superconductivity across its width.

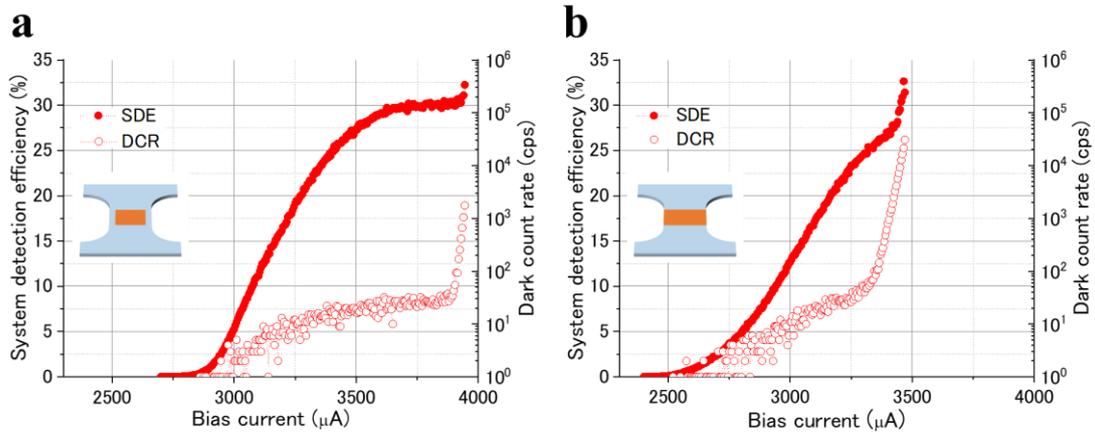

**Fig. 4 | Bias current dependences of the SDE and DCR of the SWSPD devices that were irradiated using the Ar ion beam. a**, SDE (red filled circles) and DCR (red open circles) curves of the SWSPD with the HCCB structure when measured at a temperature of 0.76 K. The wide strip was 20 μm wide, including an 18-μm-wide center strip area and 1-μm-wide side strip areas. The irradiating photon wavelength was 1550 nm. The increase in the intrinsic dark count was suppressed to the higher bias current region and a long extrinsic dark count region was observed. The SDE curve achieved saturation before the intrinsic dark count increased. **b**, SDE and DCR curves of the SWSPD without the HCCB structure. The intrinsic dark count increased from a relatively low bias current and limited device operation in the high bias current region.

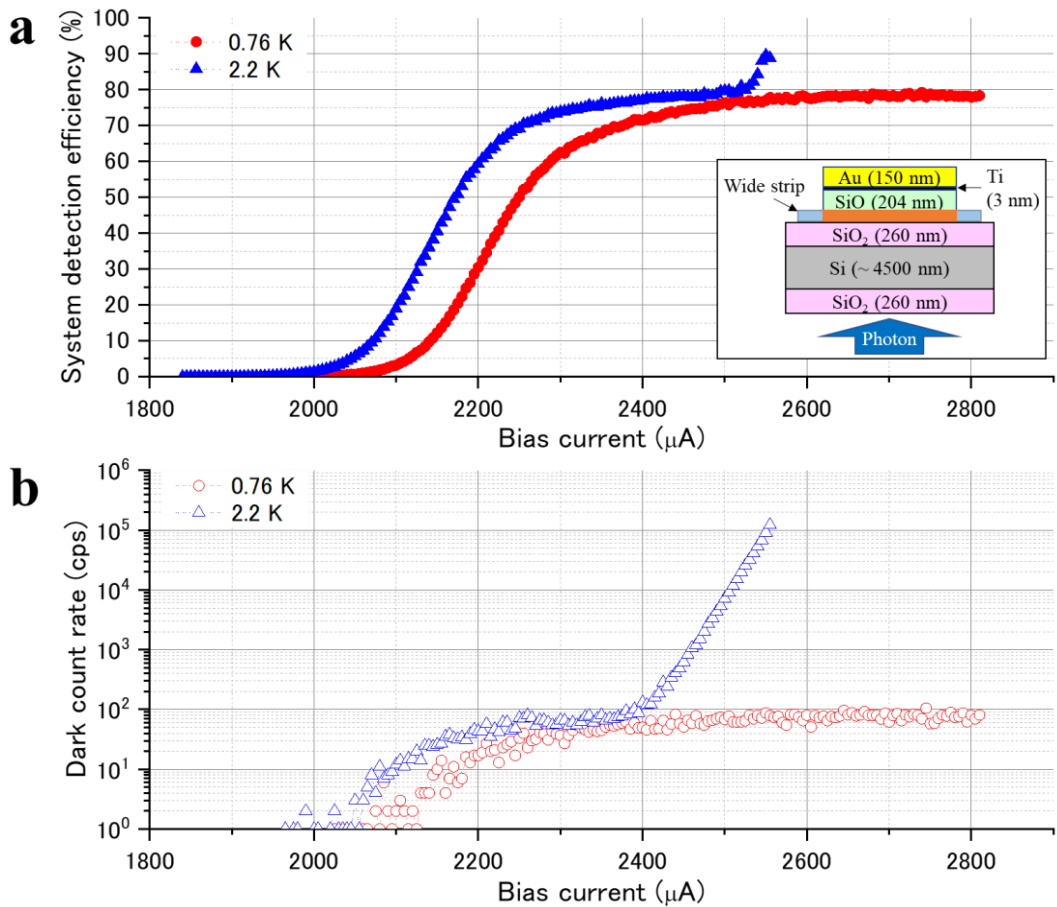

**Fig. 5 | Bias current dependences of the SDE and DCR of the SWSPD device with the HCCB structure when integrated inside the optical cavity. a**, SDE curves measured at temperatures of 0.76 K (red filled circles) and 2.2 K (blue filled triangles). The inset shows a cross-sectional view of the optical cavity structure that was designed to enhance absorption at the optical wavelength of 1550 nm. Photons with a wavelength of 1550 nm irradiated the device from the back side. **b**, DCR curves measured at temperatures of 0.76 K (red open circles) and 2.2 K (blue open triangles). The SDE of ~78% with the broad plateau region and the DCR of ~80 cps were achieved at a temperature of 0.76 K. Although the intrinsic dark count begins to increase from the lower bias current region, the SDE curve approached saturation even at the temperature of 2.2 K, and an SDE of ~76% and a DCR of less than 100 cps were achieved.

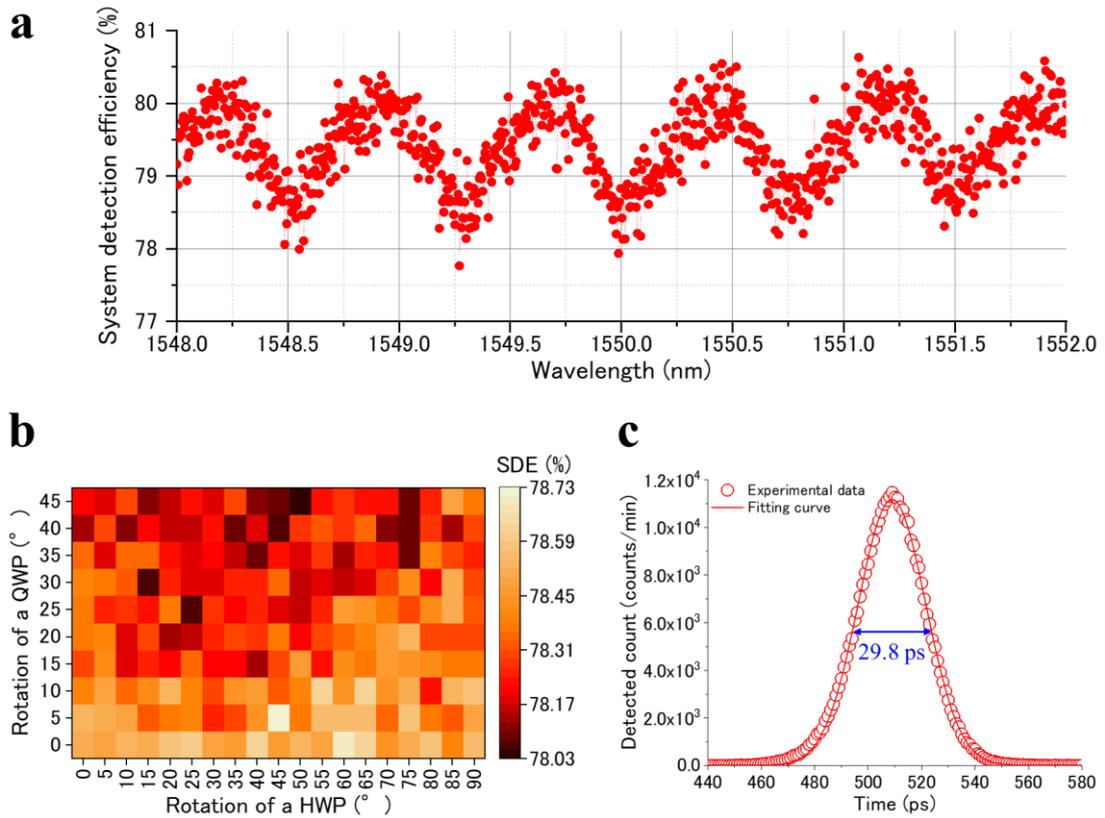

**Fig. 6 | Wavelength and polarization dependences of the SDE, and system timing jitter. a,** Optical wavelength dependence of the SDE when measured at a bias current of 2.7 mA and at a temperature of 0.76 K. The variation in the SDE is caused by optical interference induced between the back and front surfaces of the device wafer. **b**, Input photon polarization state dependence of the SDE when measured at a bias current of 2.7 mA and at a temperature of 0.76 K. The polarization state of the input photon was manipulated by rotating the half-wave plate from 0 to 90° and the quarter-wave plate from 0 to 45° in 5° steps. **c**, Timing jitter histogram measured at a bias current of 2.8 mA and a temperature of 0.76 K. The experimental data are plotted as red open circles and the red solid line shows the fitting curve plotted with the Gaussian function.

**Acknowledgements**

This work was supported by JSPS KAKENHI under Grant 22H01965 and by JST Moonshot R&D under Grant JPMJMS2066.

The authors thank Tomoya Minami for his assistance in performing device fabrication and measurements.

*Extended Data Fig. 1 | Experimental setup used for system detection efficiency and system dark count rate measurements.*

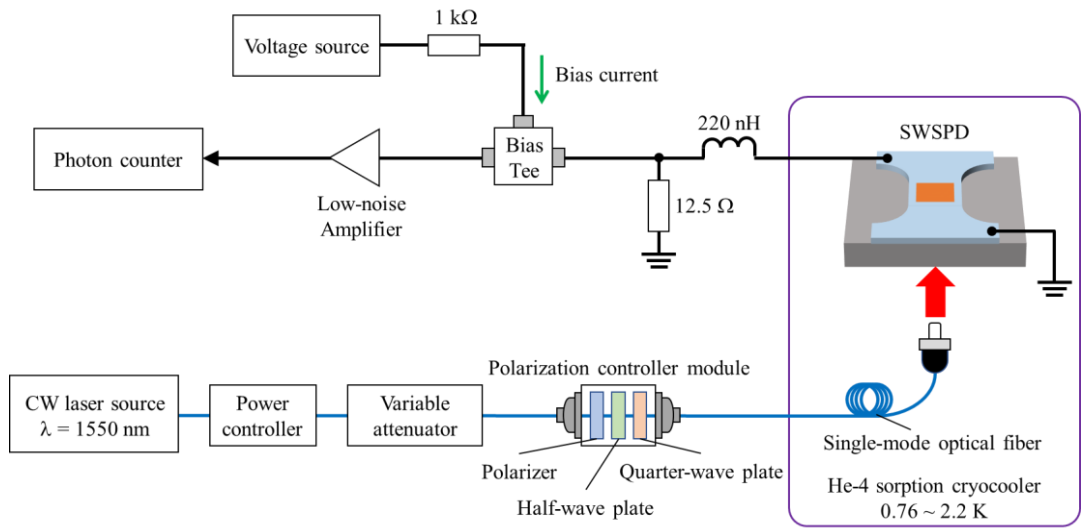

To characterize the system detection efficiency, we used a 1550 nm CW laser source as a photon source. The photons propagated through a single-mode fiber and were introduced into a He-4 sorption refrigerator after passing through an optical power controller, a variable attenuator, and a polarization control module. The output signals from the superconducting strip photon detectors were counted using a photon counter after they were amplified using a low-noise amplifier.

*Extended Data Fig. 2 | Effects of argon ion beam irradiation on the thin NbTiN superconducting strip.*

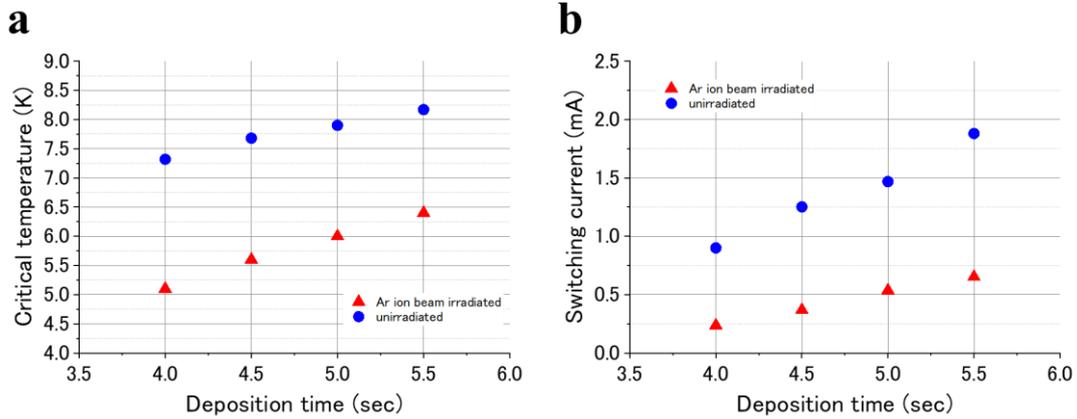

**a,** Critical temperatures and **b,** switching currents of the NbTiN superconducting strips with and without the argon (Ar) ion beam irradiation. To investigate the effects of Ar ion beam irradiation, we fabricated a NbTiN superconducting strip with a length of 20 μm and width of 10 μm on a silicon (Si) wafer with a thermally oxidized silicon dioxide ($SiO_2$) layer. To control the NbTiN strip thickness, the deposition time during the DC magnetron sputtering process was controlled within the range from 4.0 to 5.5 s, which corresponds to thicknesses of approximately 4.9 to 5.9 nm. Then, the Ar ion beam was used to irradiate the NbTiN strip for 5 s over a length of 10 μm across the entire width of the strip. All fabricated strips were cooled using a Gifford-McMahon refrigerator. The switching currents were then measured at a temperature of 2.2 K. As a result, the Ar ion beam irradiation was clearly shown to reduce the critical temperature and the switching current of the NbTiN strip. Note that the absolute values of the bias currents used in these preliminary experiments cannot be compared with those used in the main body of the article because the device design and the refrigerator system were different, and the shunt resistors and additional inductors required to stabilize device operation were not used in these experiments.

*Extended Data Fig. 3 | Experimental setup used for jitter measurement.*

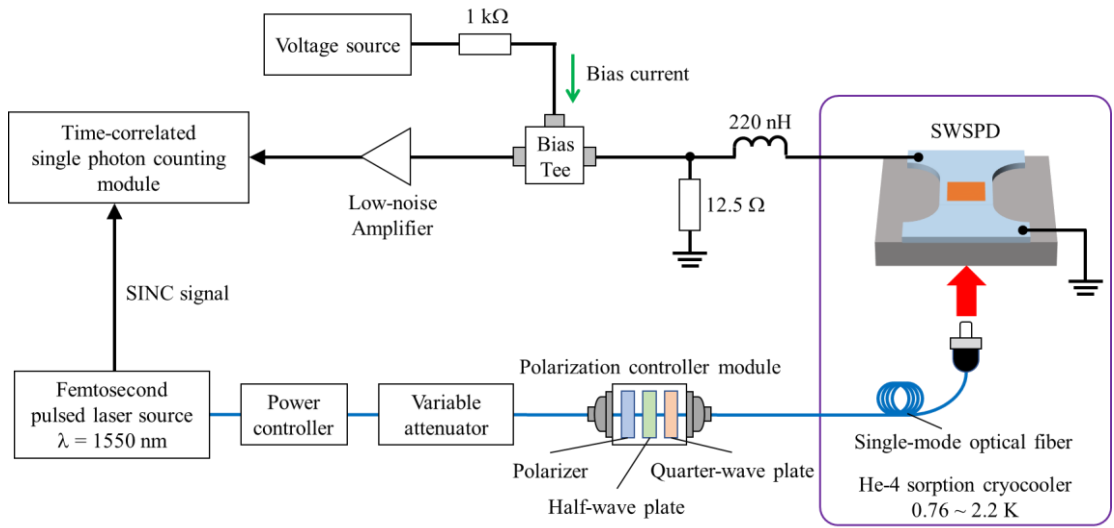

To characterize the system timing jitter, we used a 1550 nm femtosecond-pulsed laser source with pulse duration of ~100 fs and a repetition rate of 10 MHz as a photon source. The temporal correlation of the detector output signals with the synchronization signals from the laser source was measured using a time-correlated single photon counter to construct a timing histogram of the detected events.

*Extended Data Fig .4 | Output signal waveform from the SWSPD device with the HCCB structure.*

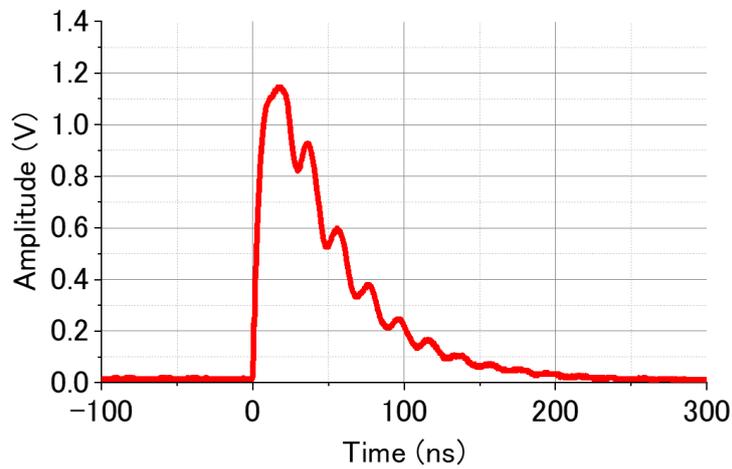

Single-shot trace of the output signal from the 20-µm-wide SWSPD device with the HCCB structure that was measured using a digital oscilloscope after being passed through a low-noise amplifier. The bias current supplied was 2.7 mA. Although ringing caused by a series inductor and a parallel resistor connected on the exterior of the refrigerator can be seen, we selected the discrimination level carefully during the system detection efficiency evaluation to ensure that the output count number was not overestimated as a result of this ringing.